\documentclass[twocolumn]{aastex631}
\usepackage{float}

\newcommand{\msun}{\mathrm{M_{\odot}}} 
\newcommand{\msunyr}{\msun\,{\rm yr}^{-1}} 

\newcommand{\mhe}{M_{\mathrm{He}}} 
\newcommand{\mch}{M_{\mathrm{Ch}}} 

\newcommand{\mdot}{\dot{M}} 

\newcommand{\Lhe}{L_{\mathrm{He}}} 

\newcommand{\iniMhe}{M^{i}_{\mathrm{He}}}
\newcommand{\iniMwd}{M^{i}_{\mathrm{WD}}}

\newcommand{\iniShe}{s^{i}_{\mathrm{c},\mathrm{He}}}

\newcommand{\kms}{\mathrm{km} \text{ } \mathrm{s^{-1}}}

\newcommand{\mesa}{{\tt\string MESA}}

\newcommand{\kB}{k_{\mathrm{B}}}
\newcommand{\NA}{N_{\mathrm{A}}}

\newcommand{\gcc}{\mathrm{g} \, \mathrm{cm}^{-3}}

\newcommand{\cp}{c_{\mathrm{p}}}

\newcommand{\She}{s_{\mathrm{c},\mathrm{He}}}

\newcommand{\vorbign}{v^{\mathrm{ign}}_{\mathrm{orb,He}}}

\newcommand{\taumdot}{\tau_{\dot{M}}}
\newcommand{\tauth}{\tau_{\mathrm{th}}}

\newcommand{\taugr}{\tau_{\mathrm{gr}}}

\newcommand{\theat}{\tau_{\mathrm{heat}}}
\newcommand{\tdyn}{\tau_{\mathrm{dyn}}}
\newcommand{\taccel}{\tau_{\mathrm{accel}}}
\newcommand{\tedd}{\tau_{\mathrm{edd}}}

\newcommand{\theatglobal}{\tau_{\mathrm{heat,global}}}

\newcommand{\Jorb}{J_{\mathrm{orb}}}

\newcommand{\Jdotgr}{\dot{J}_{\mathrm{gr}}}

\newcommand{\Porb}{P_{\mathrm{orb}}}

\newcommand{\Tc}{\ensuremath{T_{\rm c}}}

\newcommand{\Sc}{s_{c}}

\newcommand{\diff}{\mathrm{d}}

\newcommand{\cs}{c_{\mathrm{s}}}
\newcommand{\vc}{v_{\mathrm{c}}}
\newcommand{\epsnuc}{\mathcal{\epsilon}_{\mathrm{nuc}}}

\newcommand{\alphaMLT}{\alpha_{\mathrm{MLT}}}

\newcommand{\grada}{\nabla_{\mathrm{a}}}
\newcommand{\gradL}{\nabla_{\mathrm{L}}}

\newcommand{\Mcore}{M_{\mathrm{core}}}
\newcommand{\Menv}{M_{\mathrm{env}}}

\newcommand{\logTbcz}{\log_{10} \left( T_{\mathrm{bcz}} / \mathrm{K} \right) }

\newcommand{\Pbcz}{P_{\mathrm{bcz}} }
\newcommand{\Tbcz}{T_{\mathrm{bcz}} }

\begin{document}

\title{Dynamical He Flashes in Double White Dwarf Binaries}

\author[0000-0001-9195-7390]{Tin Long Sunny Wong}
\affiliation{Department of Physics, University of California, Santa Barbara, CA 93106, USA}

\author[0000-0001-8038-6836]{Lars Bildsten}
\affiliation{Department of Physics, University of California, Santa Barbara, CA 93106, USA}
\affiliation{Kavli Institute for Theoretical Physics, University of California, Santa Barbara, CA 93106, USA}

\correspondingauthor{Tin Long Sunny Wong}
\email{tinlongsunny@ucsb.edu}

\begin{abstract}

The detonation of an overlying helium layer on a $0.8-1.1\,\mathrm{M}_{\odot}$ carbon-oxygen (CO) white 
dwarf (WD) can detonate the CO WD and create a 
thermonuclear supernova (SN). 
Many authors have recently shown that when the mass of 
the He layer is low ($\lesssim 0.03\,\mathrm{M}_{\odot}$), the ashes
from its detonation minimally impact the spectra and light-curve from the CO detonation, 
allowing the explosion to appear remarkably similar to Type Ia SNe. 
These new insights motivate our investigation of dynamical He shell burning, 
and our search for a binary scenario that 
stably accumulates thermally unstable He shells in the 
$0.01-0.08\,\mathrm{M}_{\odot}$ range, thick enough to detonate, but also often thin enough for minimal impact on the observables. 
We first show that our improved non-adiabatic evolution of convective He shell burning in this shell mass range 
leads to conditions ripe for a He detonation. 
We also find that a stable mass-transfer scenario with a high entropy He WD donor of mass $0.15-0.25\,\mathrm{M}_\odot$ yields 
the He shell masses needed to achieve the double detonations. 
This scenario also predicts that the surviving He donor leaves with a space velocity consistent with the unusual runaway object, D6-2. 
We find that hot He WD donors originate in common 
envelope events when a $1.3-2.0\,\mathrm{M}_\odot$ star fills its Roche lobe at the base of the red giant branch at orbital periods of $1-10$ days with the CO WD. 

\end{abstract}


\section{Introduction} \label{sec:intro}

For decades astrophysicists have tried to answer the question -- where do type Ia supernovae (SNe Ia) come from? The broadly accepted answer is that they come from the detonation of a carbon-oxygen white dwarf (CO WD) \citep{Hoyle1960}. However, it is unclear whether SNe Ia predominantly come from explosions occurring near the Chandrasekhar mass ($\mch$), or below. 

One proposed sub-$\mch$ explosion mechanism is the double detonation scenario, where the detonation of a He shell triggers the detonation of the underlying CO core \citep[e.g.,][]{Livne1990,Livne1991,Woosley1994,GarciaSenz1999,Fink2007,Fink2010,Kromer2010,Woosley2011,Sim2012,Pakmor2012,Moll2013,ShenBildsten2015,Townsley2019,Polin2019,Gronow2020,Leung2020,Boos2021,Gronow2021}. 
A challenge to this scenario is that the burning products of the He shell detonation lead to disagreements in spectra and light curve with observations of normal SNe Ia \citep[e.g.,][]{Hoflich1996,Nugent1997}; the production of Ti, Cr and Fe group elements in the He detonation leads to line blanketing and the resulting colors are redder than observed SNe Ia \citep[e.g.,][]{Kromer2010,Woosley2011,Polin2019,Collins2022}. This can be alleviated in part by reducing the He shell mass, and, with bare sub-$\mch$ CO WDs, good agreement with observations is found \citep[e.g.,][]{Sim2010,Blondin2017,Shen2018}. With improved nucleosynthesis through the inclusion of a large nuclear network and CNO material in the He shell \citep[][]{ShenMoore2014}, \citealt{Townsley2019}, \citealt{Boos2021} and \citealt{Shen2021} find that a thin He shell double detonation ($\lesssim 0.03\,\msun$) can lead to good agreement with observations of spectroscopically normal SNe Ia 
\citep[though their thin-shell results are at variance with][]{Gronow2021,Collins2022}. 

The He detonation can be triggered by accretion stream instabilities during the dynamical phase of a double WD merger, with total accumulated He shell mass at ignition as low as $\approx 0.01 \, \msun$ \citep[e.g.,][]{Guillochon2010,Pakmor2012}. This dynamically driven double-degenerate double-detonation (D6) scenario is strongly supported by the discovery of three hypervelocity WDs with velocities $\gtrsim1000\,\kms$ \citep[][]{Shen2018_D6}. 

Alternatively, the He detonation can arise during a He shell flash where the He shell accumulates through stable mass transfer. The donor can be a nondegenerate He star, with mass transfer rates $\approx 10^{-8}\,\msunyr$ leading to a thick He shell $\approx 0.1 - 0.2 \,\msun$ \citep[e.g.,][]{Iben1991,Brooks2015,Bauer2017}, though the resulting transient better resembles peculiar SNe Ia \citep[e.g.,][]{Woosley2011,Polin2019,De2019}. 

In the AM CVn last flash scenario, the donor is a cold He WD \citep[][]{Bildsten2007,Piersanti2015,Piersanti2019}. The mass transfer rate begins high ($\mdot \gtrsim 10^{-6}\,\msunyr$), leading to weak He flashes, and decreases with time, leading to He flashes that increase in strength. The last He flash to occur is the strongest, and can potentially lead to a He detonation that results in a ``.Ia" supernova \citep[][]{Shen2010} or double-detonation SN Ia. However, \cite{Piersanti2015,Piersanti2019} suggest that the AM CVn last flash is not strong enough to become dynamical. 

We revisit He WD donors as potential progenitors of double-detonations and hypervelocity WDs like D6-2. 
Motivated by the suggestion that AM CVn binaries can be born with a wide range of entropies \citep{Deloye2007,Wong2021,vanRoestel2022,Burdge2023}, we explore a binary scenario of stable mass transfer from a high-entropy (hot) He WD onto a massive CO WD that leads to a strong, first He flash potentially developing into a detonation of the He and possibly the CO. 

We discuss the binary evolution models in Section \ref{sec:binary evolution up to ignition}. We show that high-entropy He WDs have lower peak $\mdot$ than cold He WDs, leading to accretor He shell masses at ignition comparable to, or even greater than, in the AM CVn last flash scenario \citep[][]{Bildsten2007,Piersanti2015,Piersanti2019}. In Section \ref{sec:flash}, we demonstrate that these He flashes can become dynamical and develop into a detonation, and discuss the minimum He shell mass required for such outcome. We show in Section \ref{sec:binary_scenario} that high-entropy He WDs originate in binary scenarios from unstable mass transfer with an evolved $M=1.3-2.0\,\msun$ donor near the base of the red giant branch (RGB), and a short post common envelope (CE) orbital period. We conclude in Section \ref{sec:conclusion} by discussing the open questions that remain.  


\section{Binary Evolution up to Ignition of the He Shell}
\label{sec:binary evolution up to ignition}

\subsection{Setup}

We model the mass transfer from a high-entropy He WD (donor) onto a CO WD (accretor) and the ensuing He flash on the accretor using Modules for Experiments in Stellar Astrophysics \citep[$\mesa$ version 21.12.1;][]{MESAI,MESAII,MESAIII,MESAIV,MESAV,MESAVI}. The initial WD models are constructed in version 15140. Additional He flash models in Section \ref{sec:core mass env} are run in version 22.11.1 , chosen to test the new time dependent convection capability. Our $\mesa$ input and output files are available at Zenodo (\url{https://doi.org/10.5281/zenodo.7815303}). 

We consider He WD donors with initial central entropies, $\iniShe/(\NA \kB)$, where $\NA$ is Avogadro's constant and $\kB$ is the Boltzmann constant, from $2.4$ to $4.0$ in increments of 0.1, and initial masses, $\iniMhe$, from $0.15$ to $0.25 \, \msun$ in increments of 0.01. We first evolve a $2.0$ (for $\iniMhe \leqslant 0.20 \, \msun$) or $2.5 \, \msun$ star from pre-main sequence to the formation of a He core of the desired mass (core boundary defined by hydrogen mass fraction $X=10^{-5}$), with solar metallicity \citep[$Z = 0.0142$;][]{Asplund2009} and $\mesa$'s $\tt mesa\_49.net$ network, which includes neutrons, $^{1-2}\mathrm{H}$, $^{3-4}\mathrm{He}$, $^{7}\mathrm{Li}$, $^{7,9-10}\mathrm{Be}$, $^{8}\mathrm{B}$, $^{12-13}\mathrm{C}$, $^{13-15}\mathrm{N}$, $^{14-18}\mathrm{O}$, $^{17-19}\mathrm{F}$, $^{18-22}\mathrm{Ne}$, $^{21-24}\mathrm{Na}$, $^{23-26}\mathrm{Mg}$, $^{25-27}\mathrm{Al}$, $^{27-30}\mathrm{Si}$, $^{30-31}\mathrm{P}$, $^{31-34}\mathrm{S}$ and interlinking reactions. Then we strip the envelope off using a fast wind with $\mdot$ between $10^{-8}$ and $10^{-6} \, \msunyr$, and cool the He core to the desired central entropy. 

We similarly construct CO WDs with initial masses, $\iniMwd$, of $0.9$, $1.0$ and $1.1 \, \msun$. The $0.9$ and $1.0 \, \msun$ models start with zero-age main sequence (ZAMS) masses of $5.5$ and $6.3 \, \msun$, and the $1.1 \, \msun$ model is scaled from the $1.0 \, \msun$ model after envelope stripping. The CO cores are cooled to a central temperature of $\Tc = 2 \times 10^{7} \, \mathrm{K}$. This range of CO masses is motivated by studies \citep[e.g.,][]{Sim2010,Polin2019,Boos2021,Shen2021} that predict double detonations or bare CO detonations with CO cores in this mass range yield 
spectra and light curve evolution similar to subluminous, normal, and overluminous type Ia supernovae. 

We initiate the WDs in a binary and evolve both components and orbital parameters. The initial orbital period is chosen such that the He WD comes into contact within $10^{5}$ yr and its entropy then is the same as the initial value (see Section \ref{sec:binary_scenario}). We assume fully conservative mass transfer, modeled following \citet{Kolb1990}, with orbital angular momentum loss driven solely by gravitational waves. For convergence, we set $\tt eps\_mdot\_factor = 0$ for the donor. This neglects the redistribution of energy due to mass loss as laid out in \citet{MESAV}, but has no effect on $\mdot$ since it depends on the donor's mass-radius relation (see Section \ref{sec:mdot_history}). 

We model both components as nonrotating. This could impact the stability of mass transfer, the mass transfer rate $\mdot$, and dissipation in the He shell of the accretor, all of which can influence the He shell thickness and thus strength of the He flash. 
However, as we explain here, we find any effects of rotation to be minimal. 
First, because of the larger radii of our high-entropy donors and the small mass ratio between the WDs, we find that disk accretion occurs for all runs in this study, using equation (6) of \citet{Nelemans2001}, and so the mass transfer is stable \citep{Marsh2004}. 
Second, while a fully synchronized donor will be spun up to $\approx 30 \%$ of critical rotation \citep[e.g.,][]{Bauer2021}, we found that the resulting inflated radius is similar to that of a nonrotating model with slightly higher entropy ($\Delta \iniShe \approx 0.1$). The mass transfer rate is slightly lower, but we expect only a small shift in the parameter space for a strong He flash, by $\Delta \iniShe \approx 0.1$. 
Third, \citet{Neunteufel2017} found that while angular momentum transport by the Tayler-Spruit dynamo \citep{Spruit2002} can lead to near solid-body rotation in the accretor during accretion-induced spin-up, the resulting viscous dissipation can significantly reduce the required He shell mass for ignition. 
However, \citet{Piro2008} found that viscous heating is unimportant compared to heating due to accretion for $\mdot \approx 10^{-7} \, \msunyr$ that is relevant for our work. Furthermore, we find that the enhanced Tayler-Spruit dynamo proposed by \citet{Fuller2019} reduces viscous dissipation even more considerably. 
Fourth, rotationally induced mixing between the CO core and He shell may impact ignition conditions. Studies considering hydrodynamic processes find considerable mixing. \cite{Yoon2004} find that mixing tends to stabilize He burning, while \cite{Piro2015} finds that mixing occurs earlier but at a larger depth. In contrast, \cite{Neunteufel2017}, who consider the Tayler-Spruit dynamo, find little mixing at the core-shell interface. 
Finally, tidal dissipation may also reduce the required He shell mass for ignition \citep[see][for the hydrogen, non-accreting counterpart]{Fuller2012_nova}. We defer to future studies for investigating this possibility.

During the binary run, we adopt a nuclear network that includes $^{14}\mathrm{C}$, since it participates in the $^{14}\mathrm{N}(e^{-},\,\nu)^{14}\mathrm{C}(\alpha,\,\gamma)^{18}\mathrm{O}$ (NCO) reaction chain \citep{Hashimoto1986} at densities above $1.25 \times 10^{6} \, \gcc$, and may trigger an earlier ignition for thick He shells \citep{Bauer2017}. 
We calculate the weak reaction rates linking $^{14}\mathrm{N}$ and $^{14}\mathrm{C}$ following \citet{Schwab2017}, which agree within $50\%$ for $ 5.5 \leqslant \log_{10} \left( Y_{e} \rho / \gcc \right) \leqslant 7.0 $ where $Y_{e} = 0.5$, and $ 7.0 \leqslant \log_{10} \left( T / \mathrm{K} \right) \leqslant 8.3$ with the rates provided by G. Mart\'inez-Pinedo in the $\mesa$ custom\_rates test suite, but are more finely spaced in $\log_{10} \rho$ space and so avoids interpolation issues when the rates change by orders of magnitude. The $^{14}\mathrm{C}(\alpha,\,\gamma)^{18}\mathrm{O}$ rates are from \citet{Bauer2017}.

We assume that a phase of H-rich mass transfer has already occurred prior to the He-rich phase modeled in this work \citep[e.g., extremely low-mass WDs are expected to have $\approx 10^{-3} \, \msun$ of H on their surface;][]{Istrate2016}. The H-rich mass transfer causes H flashes on the accretor, and given the short binary separation, the accretor may expand and fill its own Roche-lobe. We assume that the binary survives the H-novae by ejecting the H shells, as is found likely for a $0.2 + 1.0 \, \msun$ binary in \citet{Shen2015}. 

As He-rich mass transfer begins, the temperature in the accretor envelope rises due to compression, forming a temperature inversion, and eventually reaches ignition. We terminate the binary run once a convection zone is formed, and continue evolving the accretor through the He flash until the convection zone reaches the surface. This will be described in more detail in Section \ref{sec:flash}.


\subsection{Mass transfer history}
\label{sec:mdot_history}

High-entropy He WD donors yield a lower $\mdot$, and, when thermally unstable, a  thicker He shell at ignition than a cold He WD donor \citep[][]{Deloye2007}. By high-entropy, we mean the He WD is hotter and less dense, and has a lower degree of degeneracy. For this study, we consider He WDs with central entropy $s_{\mathrm{c},\mathrm{He}}/(\NA \kB) \gtrsim 3.0 $ high-entropy, corresponding to a cooling time $\lesssim 10^{8} \, \mathrm{yr}$ after their formation (see Section \ref{sec:binary_scenario}). With their larger radii, high-entropy He WDs reach period minimum and peak $\mdot$ at longer $\Porb$, where they have longer gravitational wave timescales, $\taugr \equiv \Jorb / \Jdotgr $ where $\Jorb$ is the orbital angular momentum and $\Jdotgr$ is the rate of angular momentum loss due to gravitational wave radiation \citep{Landau1971}. The timescale for orbital evolution is given by $\taugr$, and $\mdot \approx \mhe / \taugr $ \citep[e.g.,][]{Bauer2021}. Therefore, high-entropy He WDs have lower peak $\mdot$, as a consequence of their larger radii. There is a  donor entropy so large that the low $\mdot$ leads to accumulation of a He layer that never undergoes a flash.

\begin{figure}[hbt!]
\centering
\fig{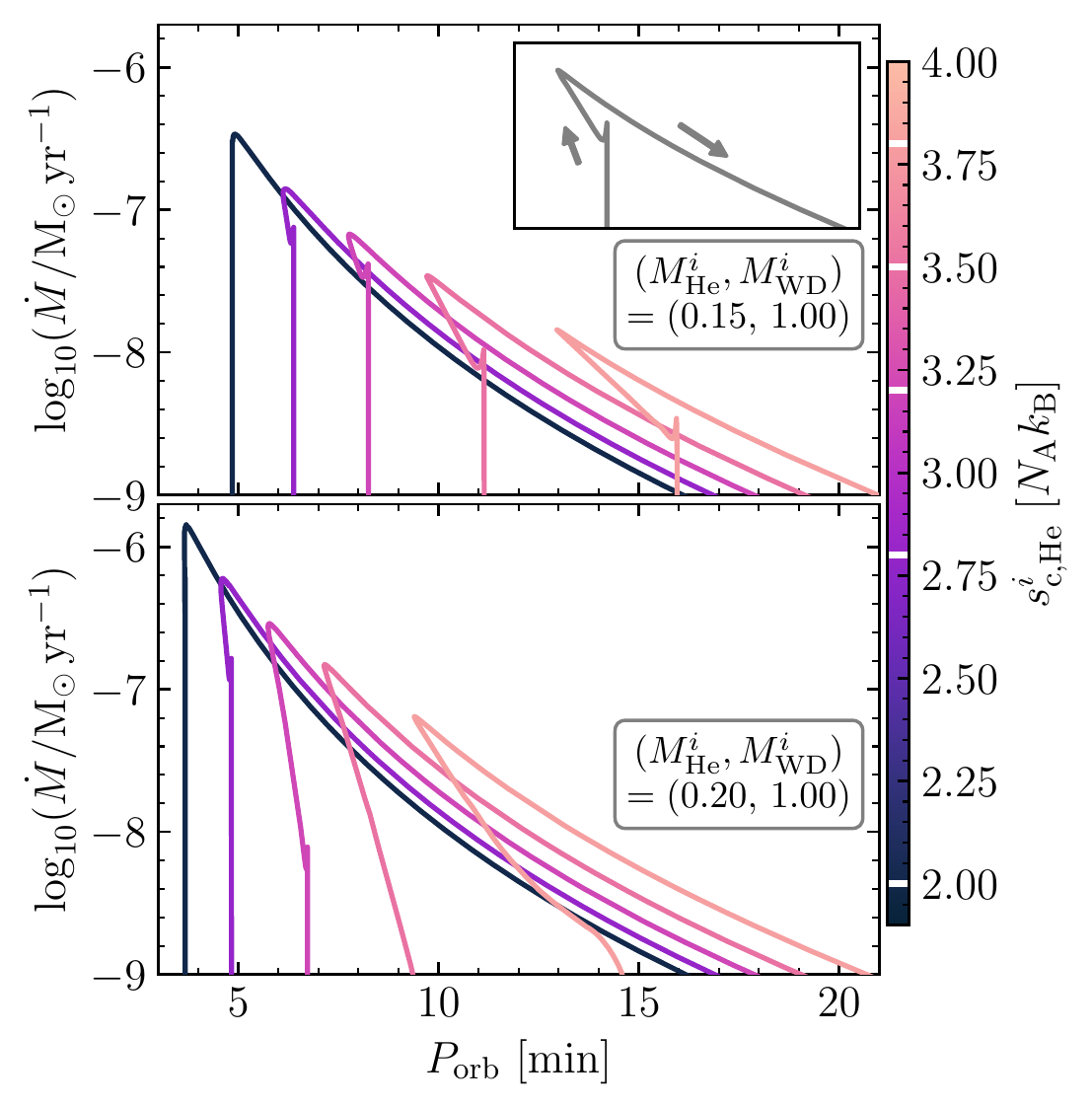}{ \linewidth }{}
\caption{ 
Mass transfer rate, $\mdot$, as a function of orbital period, $\Porb$, for He WDs with $\iniMhe = 0.15$ (top) or $0.20 \, \msun$ (bottom) of various $\iniShe$ transferring mass onto a $\iniMwd = 1.0 \, \msun$ accretor. Inset indicates the flow of time. 
\label{fig:Porb_Mdot}}
\end{figure}

Figure \ref{fig:Porb_Mdot} compares the mass transfer histories for He WD donors with $\iniMhe = 0.15, 0.20 \, \msun$, $\iniMwd = 1.0 \, \msun$, and a wide range of $\iniShe$. It illustrates that a higher-entropy He WD has a lower peak $\mdot$ and a longer period minimum. 
Higher-entropy He WDs also start with a thicker nondegenerate layer on the surface, and so come into contact at longer $\Porb$. Removal of the nondegenerate layer leads to contraction of the radius \citep{Deloye2007,Kaplan2012}, so $\Porb$ decreases and $\mdot$ gradually increases from $\lesssim 10^{-8} \, \msunyr$ to $\approx 10^{-7} \, \msunyr$. This is seen in the high-entropy models in the ``turn-on'' phase of mass transfer. 
Moreover, comparison between the top and bottom panels shows that for higher $\iniMhe$, the period minimum occurs at shorter $\Porb$, and peak $\mdot$ is higher, at fixed $\iniShe$. This is a consequence of their smaller radii. However, the $\mdot$ evolution is eventually the same regardless of $\iniMhe$, for the models with the same $\iniShe$ \citep{Deloye2007,Wong2021}. 

We do not consider initial entropies higher than $\iniShe/(\NA \kB) = 4.0$. He WDs with initial entropies below roughly this value have high enough $\mdot$ that the mass transfer timescale, $\taumdot \equiv \mhe / \mdot $, is shorter than the thermal timescale, $\tauth \equiv \int^{\mhe}_{0} \left( \cp T \diff m \right) / \Lhe $, leading to adiabatic evolution \citep{Deloye2007,Wong2021}. With $\iniShe / (\NA \kB) \gtrsim 4.0$, $\mdot$ during the ``turn-on'' phase is so low that $\taumdot > \tauth$. The donor therefore loses entropy until $\She / (\NA \kB)$ decreases down to $\approx 4.0$, where $\taumdot \lesssim \tauth$ and adiabatic evolution begins. As a result, all He WDs with $\iniShe / (\NA \kB) \gtrsim 4.0$ eventually resemble a $\iniShe / (\NA \kB) \approx 4.0$ one in $\mdot$ evolution. 


\subsection{Properties at Onset of He Flash}
\label{sec:binary_grid}

\begin{figure}[]
\fig{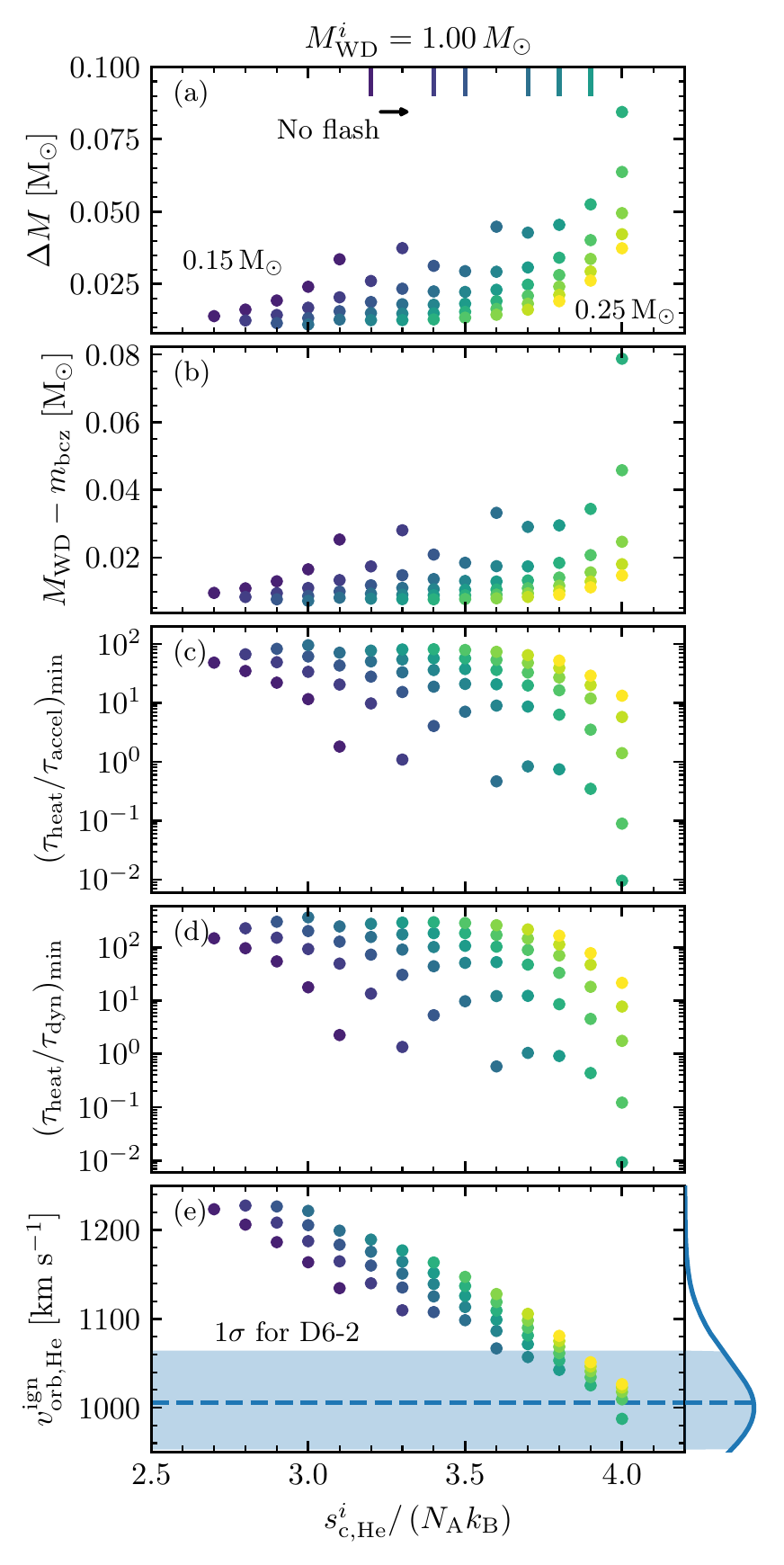}{ \linewidth }{}
\caption{ Total accumulated He shell mass at ignition, mass exterior to the base of the convection zone (BCZ), minimum $\theat/\taccel$ where convective velocity is greatest, $\theat/\tdyn$ at the BCZ, and the orbital velocity of the donor at the start of the He flash, from top to bottom. All runs start with $\iniMwd = 1.0 \, \msun$, and color-coding indicates $\iniMhe$, from $0.15 \, \msun$ for the darkest color, to $0.25 \, \msun$ for the lightest color. For each $\iniMhe$ set, we indicate the minimum $\iniShe$ above which no He flash occurs, by the lines at the top of the first panel. In the bottom panel, we show the heliocentric velocity of the hypervelocity WD D6-2 (dash-blue line) with its $1 \sigma$ uncertainty (blue region). Its posterior probability distribution \citep[][]{Bauer2021_D6} is shown on the side. 
\label{fig:grid_1p0}}
\end{figure}

Panels (a) and (e) of Figure \ref{fig:grid_1p0} show properties of the accretor and donor that are determined at He shell ignition, for our fiducial grid of $\iniMwd = 1.0 \, \msun$ models. Similar results are shown in Figure \ref{fig:other masses} for $\iniMwd=0.9,1.1\,\msun$ models. 

The total accumulated He shell masses at ignition (panel a) of our models span the range of $0.01 - 0.08 \, \msun$. These cover the range of He shell masses predicted by \cite{Bildsten2007} and \cite{Piersanti2015} for the AM CVn last flash scenario. 
Three trends can be observed. 
First, with a higher $\iniShe$ at fixed $\iniMhe$, a thicker He shell is required for ignition. This is the result of less efficient ``compressional heating'' due to the lower $\mdot$. 
The thickest He shells here are ignited with a boost from the NCO reaction chain occurring near the base of the accreted layer \citep[][]{Hashimoto1986,Bauer2017}. 
Second, no He flash occurs above a certain $\iniShe$ for each $\iniMhe$, due to the very low $\mdot$ for these models. These systems will continue to evolve as an AM CVn binary to long orbital periods \citep[e.g.,][]{Ramsay2018,Wong2021}, possibly explaining why some AM CVn donors have high-entropy \citep[][]{vanRoestel2022}.
Third, the parameter space for the same range of $\Delta M$ shifts to higher $\iniShe$ as $\iniMhe$ increases. This is because $\Delta M$ largely depends on peak $\mdot$, which is higher as $\iniMhe$ increases, and lower as $\iniShe$ increases. 

While it is unclear whether all these models can develop a He detonation, they do support a steady transverse detonation wave, especially given the inclusion of a large nuclear network and modest enrichment of CNO material \citep[][]{ShenMoore2014}. If double-detonation is successful, the ones with $\Delta M \lesssim 0.03 \, \msun $ are of interest for spectroscopically normal type Ia supernovae, while models with thicker He shells may resemble abnormal thermonuclear supernovae \citep{Polin2019,Boos2021}. 

Following (if possible) the double-detonation of the accretor and the unbinding of the binary, the donor departs at its pre-explosion orbital velocity, which is shown in panel (e) of Figure \ref{fig:grid_1p0}. The radius of the donor, and hence $\Porb$ at period minimum, increase with $\iniShe$, leading to a decrease of $\vorbign$ with $\iniShe$. Typical low-entropy ($\iniShe / (\NA \kB) \lesssim 3.0$) donors have $\vorbign \gtrsim 1100 \, \kms$, but high-entropy donors may reach $\vorbign \approx 1000 \, \kms$, becoming comparable to the heliocentric velocity of the hypervelocity WD D6-2 \citep[$1010^{+60}_{-50} \, \kms$;][]{Bauer2021_D6}. A lower $\iniMwd$ gives a lower $\vorbign$ at fixed $\iniShe$ (though the boundary for ignition may change slightly), by $\approx 40 - 50 \, \kms$ for $\iniMwd = 0.9 \, \msun$ (see Figure \ref{fig:other masses}), allowing for a larger parameter space for matching D6-2. Regardless of $\iniMwd$, our models confirm the analysis by \cite{Bauer2021_D6} that D6-2 could be a former He WD donor\footnote{Alternatively, D6-2 could be a former He star donor \citep[][]{Neunteufel2021}, or former CO WD accretor with a hybrid He/CO donor \citep{Pakmor2021}.} where a double-detonation may have happened, and our results furthermore suggest that D6-2 must have been high-entropy.

\section{The He Shell Flash}
\label{sec:flash}

Important parameters adopted during the He flash are as follows. 
First, the accretor's nuclear network is expanded to include 
neutrons, $^{1}\mathrm{H}$, $^{4}\mathrm{He}$, $^{11}\mathrm{B}$, $^{12-14}\mathrm{C}$, $^{13-15}\mathrm{N}$, $^{14-18}\mathrm{O}$, $^{17-19}\mathrm{F}$, $^{18-22}\mathrm{Ne}$, $^{21-23}\mathrm{Na}$, $^{22-26}\mathrm{Mg}$, $^{25-27}\mathrm{Al}$, $^{27-30}\mathrm{Si}$, $^{29-31}\mathrm{P}$, $^{31-34}\mathrm{S}$, 
$^{33-35}\mathrm{Cl}$, $^{36-39}\mathrm{Ar}$, $^{39}\mathrm{K}$, $^{40}\mathrm{Ca}$, $^{43}\mathrm{Sc}$, $^{44}\mathrm{Ti}$, $^{47}\mathrm{V}$, $^{48}\mathrm{Cr}$, $^{51}\mathrm{Mn}$, $^{52,56}\mathrm{Fe}$, $^{55}\mathrm{Co}$, and $^{55,56,58-59}\mathrm{Ni}$. 
This gives a network that encompasses the 55-isotope network adopted by \citet{Townsley2019} for accurate energy release. In particular, the reaction $^{12}\mathrm{C}(p,\,\gamma)^{13}\mathrm{N}(\alpha,p)^{16}\mathrm{O}$ yields significant energy release at temperatures above $10^{9} \, \mathrm{K}$ \citep{Shen2009,ShenMoore2014}, and plays an important role in some of our models. 
Second, we adopt the Cox formulation of the mixing length theory \citep[MLT;][]{Cox_and_Giuli1968}, and a mixing length parameter $\alphaMLT = 2$. We adopt the Ledoux criterion, and include semiconvective mixing \citep[with an efficiency $\alpha_{\mathrm{semi}} = 1$; ][]{Langer1985} and thermohaline mixing \citep[with an efficiency of 1; ][]{BGS2013}. 
Third, we relax the tolerances to 
\\
\indent \indent \verb| gold2_tol_residual_norm3 = 1d-6|
\\
\indent \indent \verb| gold2_tol_max_residual3 = 1d-3 |,
\\
and upon two consecutive retries, we temporarily set $\tt use\_dPrad\_dm\_form\_of\_T\_gradient\_eqn = .true.$\footnote{\url{https://docs.mesastar.org/en/release-r22.05.1/reference/controls.html\#use-dprad-dm-form-of-t-gradient-eqn}} which often aids solver convergence, particularly at the top boundary of the convection zone.

\subsection{Fiducial grid}
\label{sec:flash fiducial}


\begin{figure*}[h!]
\gridline{
\fig{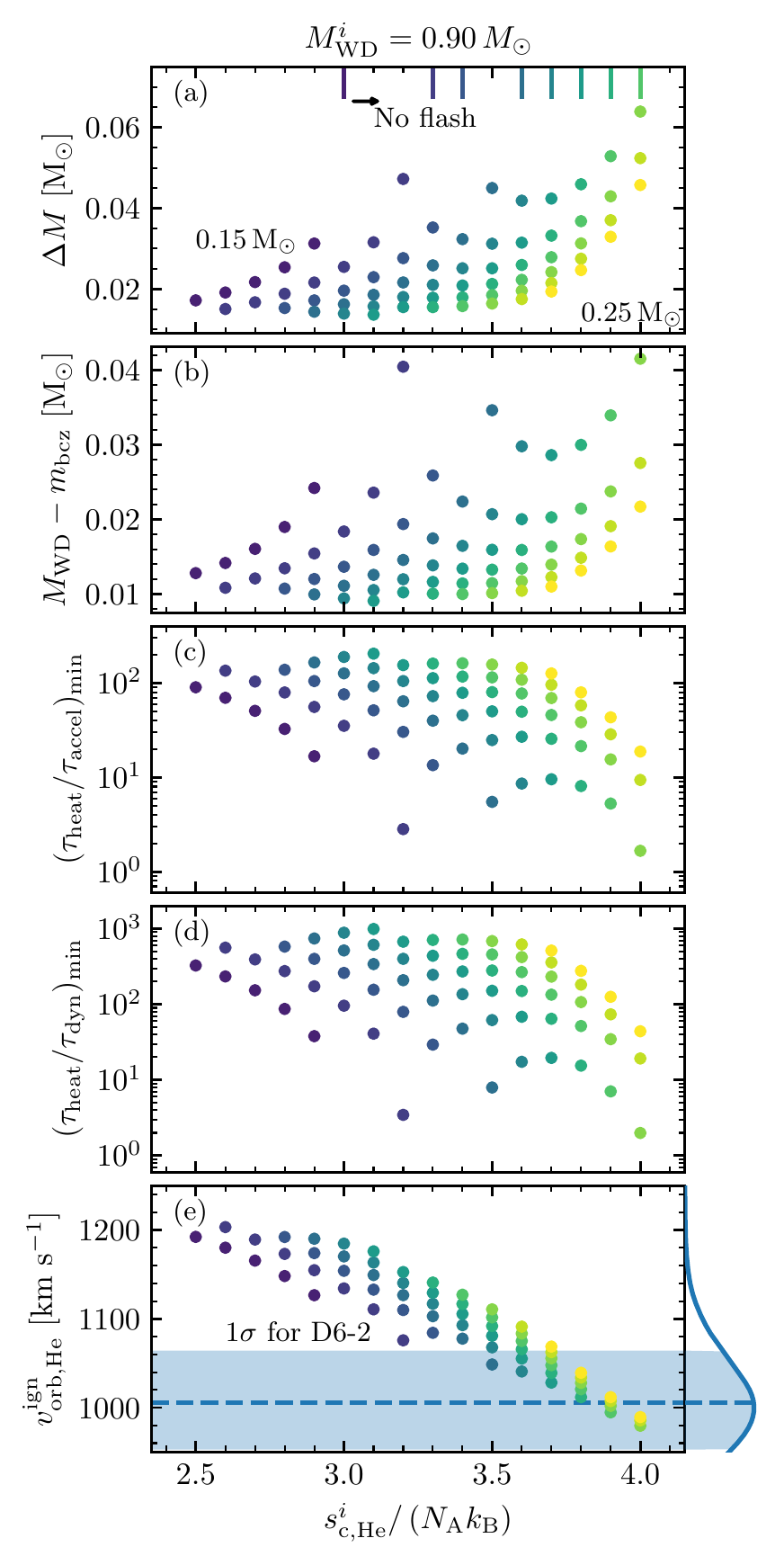}{0.5\textwidth}{(i)}
\fig{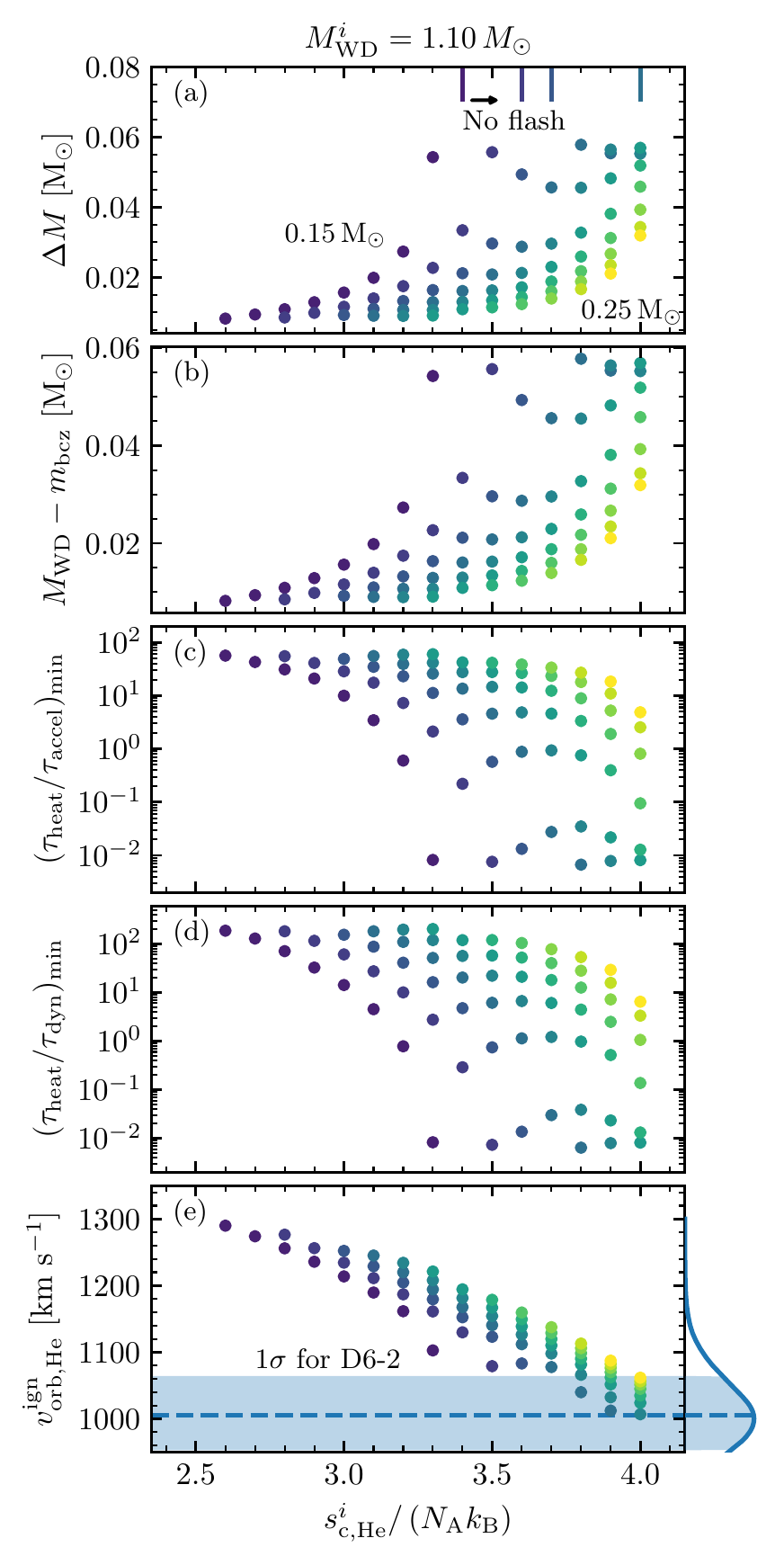}{0.5\textwidth}{(ii)}
          }
\caption{ Same as Figure \ref{fig:grid_1p0}, but with $\iniMwd=0.9,1.1\,\msun$ for the left and right panels respectively. 
\label{fig:other masses}
}
\end{figure*}


Due to heat transport away from the temperature peak during the He shell's accumulation, ignition occurs above the base of the accreted layer by $\approx 0.004 - 0.025 \, \msun$. This is illustrated by panel (b) of Figure \ref{fig:grid_1p0}, which shows the mass exterior to the base of the convection zone (BCZ) at $\logTbcz \gtrsim 8.3 $. This is slightly larger than when the convection zone first appears, because initially the inner convective boundary moves inwards until $\logTbcz \gtrsim 8.3 $. 
We do not include mixing beyond the convective boundary via overshooting, nor the convective premixing or predictive mixing schemes for determining the convective boundary \citep[e.g., ][]{MESAIV,MESAV}. 
However, any mixing beyond the convective boundary would move the BCZ further inwards, creating a more explosive outcome. In Section \ref{sec:core mass env}, we artificially induce ignition at the base of the accreted layer. Furthermore, in Section \ref{sec:other uncertainties} we show the effects of adopting a different accretor $\Tc$ and conductive opacity, both of which influence conditions at the BCZ. 


Three timescales affect the outcome of the He flash. The first is the dynamical timescale, 
\begin{equation}
    \tdyn = \frac{ H }{ \cs } ,
\end{equation}
where $H$ is the pressure scale height and $\cs$ is the sound speed.
The second is the local heating timescale, at which temperature increases due to burning, 
\begin{equation}
    \theat = \frac{ \cp T }{ \epsnuc }, 
\end{equation}
where $\cp$ is the specific heat capacity and $\epsnuc$ is the nuclear energy generation rate. This is smaller than the global heating timescale over the convective envelope, $\theatglobal = \int ( \cp T \, \diff m ) / \int ( \epsnuc \, \diff m )$, which is the time to heat the entire convection zone \citep[][]{Shen2009}. 
The third is the convective acceleration timescale, at which convective velocity, $\vc$, varies \citep{MESAVI} and which is given by, in $\mesa$ default parameters, 
\begin{equation}
    \taccel = \frac{ 3 H }{ \sqrt{ 2 \cp T \grada \left( \nabla - \gradL \right) } }, 
\end{equation}
where $\nabla$ is the temperature gradient, $\grada$ is the adiabatic gradient, and $\gradL$ is the Ledoux gradient. In steady state, $\taccel$ is $3/(2 \alphaMLT)$ times the eddy turnover timescale, 
\begin{equation}
    \tedd = \frac{ H }{ \vc }. 
\end{equation}

When $\theat \lesssim \taccel $, the standard assumption that convection is in steady state becomes dubious, since the temperature rises faster than convection can respond. Instead, convection is expected to freeze out \citep[e.g., ][]{Woosley2011,MESAVI}. When time-dependent convection (TDC) is applied, heat is more strongly trapped at the BCZ, leading to strong superadiabaticity and a higher peak $\Tbcz$ \citep[see ][]{MESAVI}. 
\cite{Woosley1994} also argue that convection breaks down when $\theat\lesssim\tedd$. A subsonic, turbulence-dominated deflagration results and, although not well-studied for He, may transition into a detonation \citep[e.g.,][]{Shen2010}. 

When $\theat \lesssim \tdyn$ locally, an overpressure develops over the scale height and a detonation is very likely. Moreover, \cite{ShenMoore2014} show that, if a large nuclear net and CNO isotopes are included, the detonation may well initiate in a hotspot that is small (required to be at least $\approx 3 \times 10^{6} \, \mathrm{cm}$ for an isobaric hotspot with central temperature $10^{9}\,\mathrm{K}$ and central density $10^{5}\,\gcc$) compared to the scale height of the convection zone ($\approx$ few $10^{7}-10^{8}\,\mathrm{cm}$). In this case, the local heating timescale should be compared to the sound-crossing time over the dimension of the hotspot, which makes a detonation even more likely \citep[][]{ShenMoore2014}. In addition, many He flashes realized in this work are ignited above the base of the accreted layer and mixing beyond the convective boundary can induce a stronger He flash. For these two reasons, we consider $\theat \lesssim 100 \, \tdyn$ of interest, as long as the envelope can sustain a steady transverse detontation wave. 
We note that a hydrodynamical approach is more appropriate in the limit $\theat\lesssim\tdyn$, as is done by \cite{Woosley2011} in 1D, but we continue to adopt a hydrostatic approach for numerical convenience and to approximate $\theat / \tdyn$ of our models. 
Furthermore, when $\theat\approx\tdyn$, $\theat$ already approaches $\approx 0.1 $ times $ \pi r_{\mathrm{bcz}} / \cs$, i.e., the time for sound waves to communicate over a shell of radius $r_{\mathrm{bcz}}$. In other words, multiple points at the BCZ may initiate an ignition \citep[e.g.,][]{Woosley2011}. 
Future three-dimensional simulations similar to \citet{Zingale2013}, \citet{Jacobs2016}, and \citet{Glasner2018} may further inform the exact conditions of the initiation of a detonation. 

The minimum values of the ratios $\theat/\taccel$ and $\theat / \tdyn$ are compared in panels (c) and (d) of Figure \ref{fig:grid_1p0}. 
A thicker He shell leads to a higher $\Pbcz$ for hydrostatic balance, and a higher peak $\Tbcz$. These both lead to stronger nuclear burning, and hence lower $\theat$. Therefore, $\theat/\taccel$ and $\theat / \tdyn$ both decrease with $\Delta M$, and hence $\iniShe$. 
Some of our models, with total accumulated He masses of $\gtrsim 0.03 \, \msun$, can reach $\theat / \tdyn \lesssim 10 $. This suggests that high-entropy He WD are a viable channel for He detonations and related transients.


\subsection{Different accretor mass}
\label{sec:different accretor mass}

In Figure \ref{fig:other masses}, we show the results for an initially $\iniMwd=0.9,1.1\,\msun$ accretor. Both show a similar range of total accumulated He shell mass at ignition as the fiducial $1.0\,\msun$ grid, but as $\iniMwd$ increases, the minimum ratio between $\theat$ and $\tdyn$, and $\theat$ and $\taccel$ decrease. This results from the increasing density at the He base as $\iniMwd$ increases. 
As $\iniMwd$ increases, so does the surface gravity and density at the base of the He shell. As a result, for the same $\iniMhe$ and $\iniShe$, the total accumulated He shell mass at ignition, minimum ratio between $\theat$ and $\tdyn$, and $\theat$ and $\taccel$, decrease with $\iniMwd$. In other words, for a given $\iniMhe$, in order to achieve a dynamical He flash, a higher $\iniMwd$ requires a slightly lower $\iniShe$. 
Also due to the increasing density at the He shell base with $\iniMwd$, none of the $0.9\,\msun$ models show ignition boosted by the NCO reaction chain, but highest entropy models with $\iniMwd=1.0,1.1\,\msun$ do, as the density at the He base reaches the critical density $1.25\times10^{6}\,\gcc$ \citep[][]{Bauer2017}.


\subsection{Other variables}
\label{sec:other uncertainties}

In this work we include the correction by \cite{Blouin2020} to the electron conductive opacities, which results in a lower opacity in the He envelope of the accretor. Due to the faster transport of heat away from the envelope, a higher He shell mass is required for ignition. However, the \cite{Blouin2020} correction, while accurate at moderate degeneracy, may not be appropriate at strong degeneracy \citep[][]{Cassisi2021}. We re-ran our $\iniMhe=0.15\,\msun$, $\iniShe/(\NA \kB)=3.1$ binary applying the damping factors proposed by \cite{Cassisi2021} to the accretor. With a `weak (strong) damping', the total accumulated He mass at ignition is reduced from $0.034\,\msun$ to $0.032(0.027)\,\msun$, while the mass enclosed by the BCZ is increased from $1.008\,\msun$ to $1.012(1.012)\,\msun$. Both result in a slightly weaker He flash, with the minimum $\theat/\tdyn$ increased from 2.2 to 6.5(23.8). Nevertheless, the uncertainty in the electron conductive opacities plays a minor role in our simulations, compared to, e.g., the mass transfer history. 

While we fix the initial center temperature of the accretor to $\Tc=2\times10^{7}\,\mathrm{K}$, we test the effects of different choices of $\Tc$ by re-running the 
$\iniMhe=0.15\,\msun$, $\iniShe/(\NA \kB)=3.0$ binary. The fiducial run has a total accumulated He mass of $0.024\,\msun$ and a mass enclosed by the BCZ of $1.008\,\msun$, which are changed to $0.027,0.022,0.019\,\msun$ and $1.010,1.005,1.002\,\msun$ for $\log_{10}(\Tc/\mathrm{K})=7.0,7.5,7.7$ respectively. In other words, a hotter initial accretor results in a lower accumulated He mass at ignition and ignition closer to the center \citep[e.g., ][]{Woosley2011}. These opposing effects largely cancel and result in a similar minimum $\theat/\tdyn$, with 17.9 for the fiducial and 15.5, 19.8, and 22.3 for $\log_{10}(\Tc/\mathrm{K})=7.0,7.5,7.7$ respectively. 


\subsection{Envelope mass condition for dynamical flash}
\label{sec:core mass env}

\begin{figure}[]
\fig{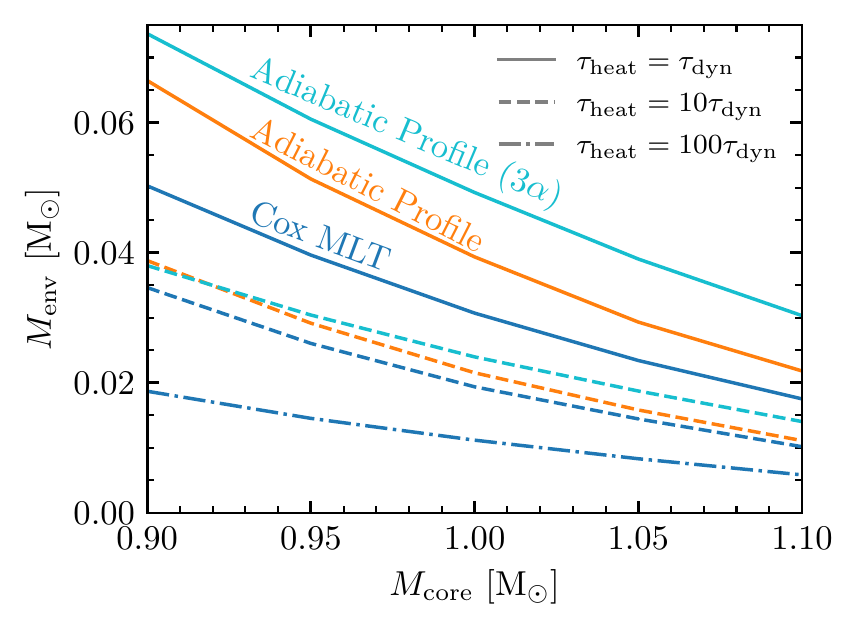}{ \linewidth }{}
\caption{ He envelope mass required such that $\tdyn/\theat$ at the BCZ reaches 1 (solid lines), 10 (dashed lines), and 100 (dot-dashed; only for Cox MLT) for a given mass enclosed by the BCZ. Dark blue, orange and light blue lines correspond to Cox MLT, adiabatic profile, and adiabatic profile with only triple-alpha burning. 
\label{fig:core_env}}
\end{figure}

Our models differ from those in \citet{Shen2009} because ours allow superadiabaticity in the convective zone and use a large nuclear net. We now explore the impacts of these two choices on the envelope mass required for a dynamical flash. 

We construct He flash models with different combinations of core mass $\Mcore$ and envelope mass $\Menv$ as follows. We first scale the $1.0\,\msun$ CO WD ($X(^{16}\mathrm{O})\approx 0.61,X(^{12}\mathrm{C})\approx 0.37,X(^{22}\mathrm{Ne})\approx 0.02$) to the desired $\Mcore$. Then we accrete material similar in composition to the $0.15\,\msun$ He WD ($X(^{4}\mathrm{He})\approx0.986$, $X(^{14}\mathrm{N})\approx0.0088$, where the progenitor star has $Z=0.0142$), onto the CO WD until $\Menv$ is reached, at $\mdot=10^{-8}\,\msunyr$ so that the He does not ignite. The CO WD is allowed to cool until $\Tc=2\times10^{7}\,\mathrm{K}$. Finally, the envelope is artificially heated at its base until a convection zone appears. In other words, unlike in the binary evolution, the BCZ is located at the base of the accreted material. 

We run grids of models with 3 different treatments of convection: (1) with Cox MLT allowing superadiabaticity, which is the same as in Section \ref{sec:flash fiducial}; 
(2) forcing an adiabatic profile in the convection zone;
and (3) forcing an adiabatic profile, accounting for only the triple-alpha reaction and not allowing compositional changes so as to simulate a nearly pure He envelope, as is assumed in \citet{Shen2009}. Adiabatic convection is enforced by the $\verb|MLT++|$ capacity \citep[][]{MESAII}, via the $\mesa$ controls:
\\
\indent \indent \verb| okay_to_reduce_gradT_excess = .true.|
\\
\indent \indent \verb| gradT_excess_lambda1 = -1 |
\\
\indent \indent \verb| gradT_excess_max_logT = 12 |,
\\
which, together, ensure that superadiabaticity in the convection zone is fully reduced.

From the grids of models, we interpolate in $\Mcore$ and $\Menv$ to find where $\theat=\tdyn$, $\theat=10\,\tdyn$, and $\theat=100\,\tdyn$ (for Cox MLT only). These are shown in Figure \ref{fig:core_env}. For a given $\Mcore$, 
the adiabatic profile with only triple-alpha burning requires the thickest $\Menv$, followed in order by the adiabatic profile and Cox MLT. 
The difference between the first two reflects the importance of including a large nuclear net, in particular the reaction $^{12}\mathrm{C}(p,\,\gamma)^{13}\mathrm{N}(\alpha,p)^{16}\mathrm{O}$ \citep[][]{Shen2009,ShenMoore2014}, though this is slightly metallicity-dependent since the protons are produced from reactions like $^{14}\mathrm{N}(\alpha,\gamma)^{18}\mathrm{F}(\alpha,\mathrm{p})^{21}\mathrm{Ne}$ \citep[][]{Shen2009}. However, we varied the metallicity of the $\Mcore=1.0\,\msun$, $\log_{10}(\Menv/\msun)=-1.5$ model, and found that the minimum $\theat/\tdyn$ changes from $1.45$ at $0.1\,Z_{\odot}$, to $0.86$ at $Z_{\odot}$ (our fiducial), and to $0.77$ at $2\,Z_{\odot}$, so the uncertainty resulting from varying metallicity plays a small role.  
The difference between the adiabatic profile and Cox MLT arises because, for a given $\Mcore$ and $\Menv$, superadiabaticity as allowed by Cox MLT results in a higher peak $\Tbcz$, which in turn reduces $\theat$. 
Finally, we further run a grid of models with TDC (again allowing superadiabaticity), which agrees well with Cox MLT for less dynamical flashes. With more dynamical flashes, because the inequality $\theat\lesssim\taccel$ strengthens, TDC exhibits stronger heat-trapping at the BCZ \citep[see Section 3.6 of ][for more details]{MESAVI}. This results in a larger superadiabaticity and a larger peak $\Tbcz$. However, the reduction in $\Menv$ required for $\theat=\tdyn$ is $\lesssim 5\%$. 

In agreement with \cite{Woosley2011}, we find that the $^{12}\mathrm{C}(p,\,\gamma)^{13}\mathrm{N}(\alpha,p)^{16}\mathrm{O}$ reaction reduces the minimum $\Menv$ required for a dynamical He flash. Our Cox MLT, $\theat = \tdyn$ line agrees well with their ``hot'' line (see their Figure 19) at $\Mcore\approx0.9\,\msun$, but is slightly lower by $\approx30\%$ at $\approx1.1\,\msun$. The reason could be that they define $\theat$ as the time to run away to $1.2\times10^{9}\,\mathrm{K}$ and they have a sparser grid of models. 


\section{How to make high-entropy He WDs?}
\label{sec:binary_scenario}

Now the question is, how does one obtain He WDs that are high-entropy at contact? We address this first assuming that the He WD is formed through a CE event. 

\begin{figure*}[t!]
\centering
\fig{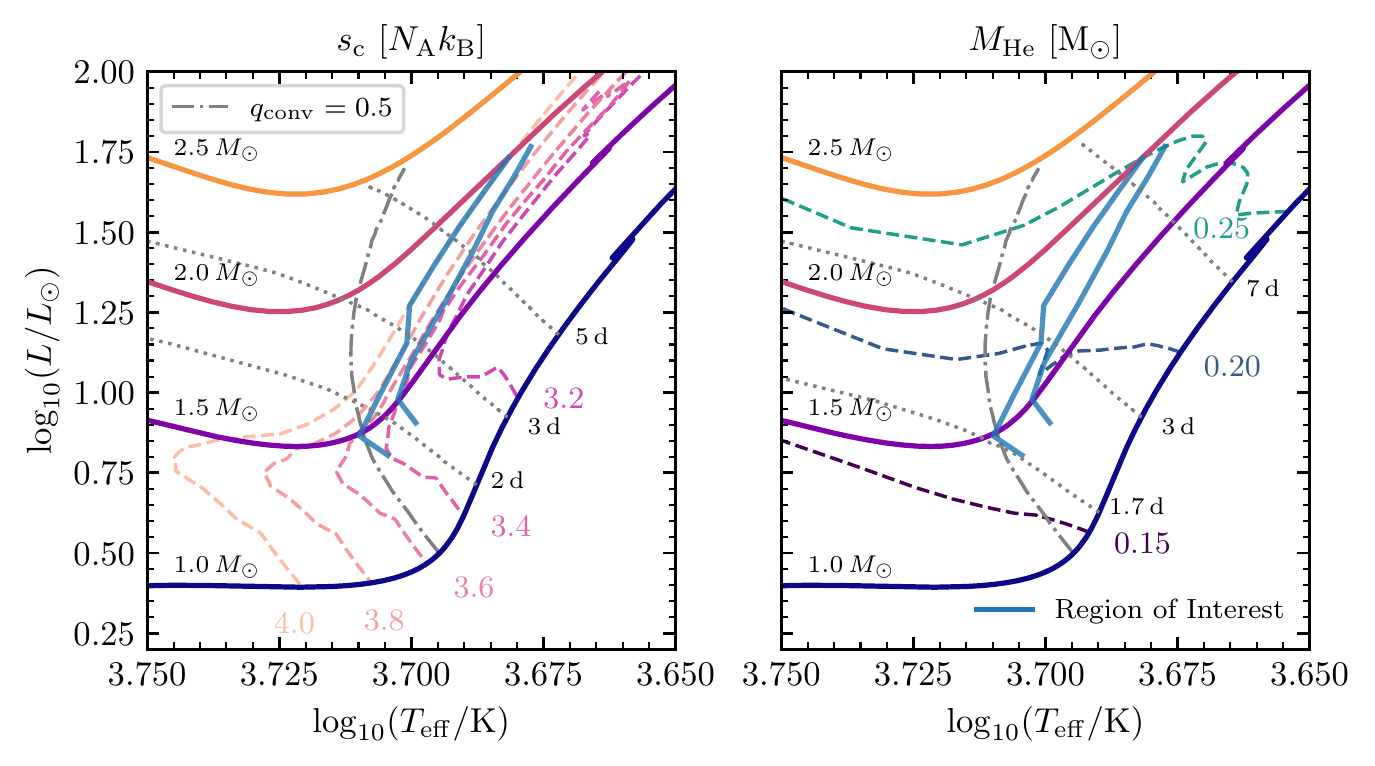}{ 0.8 \textwidth }{}
\caption{ Evolution of $1.0-2.5 \, \msun$ stars up the RGB, with contours labeling central specific entropy (left) and He core mass (right). The two blue lines bracket parameter space for which a dynamical He flash may happen once mass transfer occurs after the CE event. The convective envelope occupies half of the star's total mass to the right of the dot-dashed line. The three dotted lines label the orbital periods of 2, 3, 5 days (left), and 1.7, 3, 7 days (right) where the star fills its Roche lobe with a $1.0\,\msun$ companion. 
\label{fig:HR}}
\end{figure*}

First, the He WD has to be high-entropy at formation. Figure \ref{fig:HR} shows the evolution on the HR diagram of stars of $M=1.0-2.5\,\msun$ (without mass loss), from the start of core H burning through the RGB. Contours label $\Sc$ and $\mhe$ (where the He core boundary is defined by H mass fraction $X=0.1$ in this section) in the left and right panels respectively. As the He core mass grows while the star crosses the Hertzsprung gap (HG) and ascends the RGB, $\Sc$ drops. With higher MS progenitor mass, a given $\mhe$ is formed earlier with higher $\Sc$. As the CE event happens on short timescales, $\Sc$ of the post-CE He WD is the same as that of the pre-CE He core. 
Thus, if a CE event occurs early in the post-MS evolution (near the base of the RGB)\footnote{We assume a CE event will happen since the RGB star has a deep convective envelope ($q_{\mathrm{conv}} \gtrsim 0.5$). For more realistic conditions, see, for example, \cite{Temmink2023}.}, corresponding to an orbital period of $\approx 1 - 8 \, \mathrm{d}$ for a $1.0\,\msun$ companion, then a high-entropy He WD can be obtained at formation. 

\begin{figure}[]
\centering
\fig{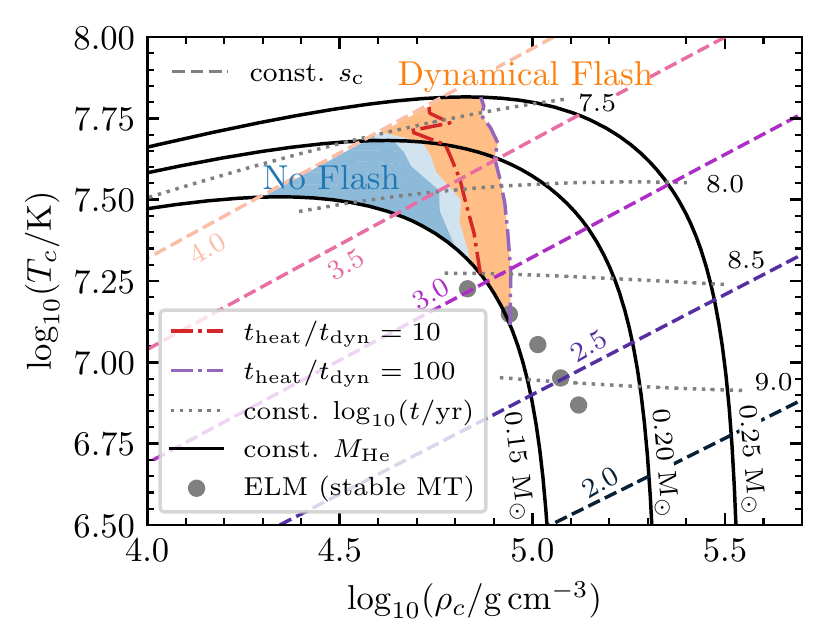}{ \linewidth }{}
\caption{ Central temperature and density trajectories of cooling He WDs of fixed masses from $0.15$ to $0.25 \, \msun$ (black solid lines). Lines of constant entropy and cooling age are in dashed and dotted respectively. Blue region gives the parameter space where no He flash occurs, orange region where a He flash is possible with $\theat\lesssim100\,\tdyn$, and light blue unknown due to our finite model grid. We also limit our models to $\iniShe/(\NA \kB) \leqslant 4.0$. Dot-dashed lines are estimates of where a He flash occurs with $\theat=10\,\tdyn$ (red) and $100\,\tdyn$ (purple). Circle markers are our ELM models following \cite{Sun2018} which descend from the stable mass transfer channel. 
\label{fig:cooling}}
\end{figure}

Second, this high-entropy He WD should not cool before coming into contact. Figure \ref{fig:cooling} shows the cooling evolution of He WDs of various masses, in central temperature and density space. Comparison between the lines of constant $\Sc$ and those of constant cooling age, shows that the high entropies required by our scenario, $\Sc/(\NA\kB)=3.0-4.0$, implies that the He WDs can only cool for $\lesssim 10^{8}\,\mathrm{yr}$ between their formation and the onset of He mass transfer. 

Third, in order to have cooled for little time before contact, the binary has to be formed at short orbital periods. For example, the gravitational-wave-induced merger timescale, which can be taken as the time before the binary comes into contact again, is $\approx10^{8}\,\mathrm{yr}$ for a $0.2\,\msun+1.0\,\msun$ binary  with an orbital period of $\approx1\,\mathrm{hr}$. 

Such short post-CE periods are favored by recent findings that the CE efficiency is low \cite[e.g.,][]{Zorotovic2010,Scherbak2023}, as well as suggestions that ELM WDs are formed at short orbital periods \citep[][]{Brown2016}. Along each evolutionary track in Figure \ref{fig:HR}, we assume that the He WD progenitor (star 1) fills its Roche lobe and undergoes a CE event, during which the companion (star 2) remains at the same mass. Assuming that the change in orbital energy during the CE event is used to eject the CE with an efficiency of $\alpha \approx 1/3$ \citep[e.g.,][]{Scherbak2023}, we solve the CE energy equation for the post-CE binary separation, $a_{f}$, 
\begin{equation}
    E_{\mathrm{bind}} = \alpha \left( \frac{GM_{1}M_{2}}{2a_{i}} - \frac{G \mhe M_{2}}{2a_{f}} \right), 
\end{equation}
\newline
where $E_{\mathrm{bind}}$ is the binding energy of the envelope obtained from $\mesa$ accounting for recombination energy, and the pre-CE binary separation, $a_{i}$, is given by the RLOF condition for star 1 \citep[][]{Eggleton1983}
\begin{equation}
    a_{i}= R_{1} \frac{0.6q^{2/3}_{1} + \ln\left(1+q^{1/3}_{1}\right)}{0.49q^{2/3}_{1}}, 
\end{equation}
where $q_{1}=M_{1}/M_{2}$. 

A low CE parameter e.g., $\alpha=1/3$ favors formation of He WD binaries at ultrashort periods $\Porb\ll1\,\mathrm{hr}$, which have very short gravitational wave merger timescales. 
For systems with $\mhe\gtrsim0.25\,\msun$ and $M\lesssim1.5\,\msun$, there is sufficient time for the newly formed He WD to cool, but they do not lead to dynamical flashes. 
In contrast, systems that do lead to dynamical flashes, marked as the region of interest in Figure \ref{fig:HR}, will remain at the same entropy. 
However, the post-CE binary may be so compact that the newly-formed He WD immediately fills its Roche lobe \citep[e.g.,][]{Deloye2007}. We remain agnostic as to whether this leads to a merger outcome or that a transition to stable mass transfer is possible. If the latter case does occur, our work suggests that a CE event occurring at the base of the RGB favors the formation of a high-entropy He WD donor amenable to a dynamical He flash from later mass transfer. 

In the stable mass transfer channel for ELM WD formation, a pre-ELM WD with mass $\gtrsim0.15\,\msun$ becomes detached at $\Porb\gtrsim5\,\mathrm{hr}$ \citep[e.g., ][]{Sun2018}. 
Although these models have thick H shells and undergo stable H burning which maintains a warm He core \citep[][]{Kaplan2012}, the He WD will still have cooled for too long and become low-entropy before coming into contact again. This is show by the grey dots in Figure \ref{fig:cooling}, where the models are made following \cite{Sun2018}. 
Given this, we suggest that a high-entropy He WD has to descend from the CE channel with a short post-CE orbital period $\Porb\lesssim1\,\mathrm{hr}$. 
However, we note that an appreciable number of observed ELM WD binaries with $\mhe\lesssim0.2\,\msun$ have shorter $\Porb$ than predicted by the stable mass transfer channel \citep[][]{Li2019}, so it is possible the stable mass transfer channel may still produce high-entropy He WDs that are of interest to this work.

\section{Conclusion}
\label{sec:conclusion}

We have shown that mass transfer from a high-entropy He WD onto a massive CO WD can lead to a strong, first He flash. 
With higher donor entropy, the peak $\mdot$ decreases, leading to a larger total accumulated He shell mass at ignition (see Section \ref{sec:binary evolution up to ignition}). For an initially $1.0\,\msun$ accretor, the explored range of total accumulated He shell mass at ignition spans from $0.01$ to $0.08\,\msun$. 
By including CNO isotopes and a large nuclear network accounting for the reaction chain $^{12}\mathrm{C}(p,\,\gamma)^{13}\mathrm{N}(\alpha,p)^{16}\mathrm{O}$ \citep[][]{ShenMoore2014}, and by allowing superadiabaticity in the convection zone, we show that the resulting He flash can become dynamical 
(see Section \ref{sec:flash}). 

Pending a successful double detonation, this scenario can explain some SNe Ia. For some thin-shell ($\lesssim0.03\,\msun$) models in our simulations, the resulting transient may be a spectroscopically normal SNe Ia, and thick-shell may produce peculiar SNe Ia \citep[e.g.,][]{Polin2019,Townsley2019,Boos2021,Shen2021}. 
Furthermore, our results provide a good match to the velocity of the hypervelocity WD D6-2 \citep[][]{Shen2018_D6,Bauer2021_D6}, and suggest that D6-2 must have been high-entropy. We plan to study the impact of SN ejecta on high-entropy He WD donors like D6-2 in the near future, similarly to \cite{Bauer2019}. 

For non-dynamical He flashes that cause expansion of the accretor to Roche-lobe overflow, \cite{Shen2015} has raised the possibility of a subsequent merger. Should this be the case, then the only surviving systems that continue to evolve to longer periods as AM CVn binaries \citep[e.g.][]{vanRoestel2022}, are those with initial entropies so large that no flash ever occurs. 

In Section \ref{sec:binary_scenario}, we argue that these high-entropy He WDs that have a dynamical flash must result from a CE event. The unstable mass transfer begins near the base of the RGB when the donor progenitor is still high-entropy, and ends with short orbital periods such that the newly formed He WD cannot cool before coming into contact again. Main sequence stars with masses $\approx 1.3-2.0\,\msun$, an unseen companion with mass $\approx 1.0\,\msun$, and orbital periods between 1 and 8 days are good candidates for producing the high-entropy He WDs of interest here. The short lifetimes ($1.1-3.5$ Gyr) of these main sequence stars may explain SNe Ia in a younger population. 


\begin{acknowledgements}

We thank the referee for their constructive suggestions that have greatly improved our manuscript. 
We thank Evan Bauer, Adam Jermyn, Jared Goldberg, and Will Schultz for helpful conversations about running $\mesa$ and Evan Bauer in addition for sharing data of D6-2. 
We are grateful to Abigail Polin for helpful conversations about He detonation and thermonuclear supernovae, and Ken Shen in addition for helpful comments on an earlier draft. 
We thank Meng Sun for sharing her $\mesa$ inlists for modeling ELM WD formation. 
This work was supported, in part, by the National Science Foundation through grant PHY-1748958, and by 
the Gordon and Betty Moore Foundation through grant GBMF5076. 
Use was made of computational facilities purchased with funds from the National Science Foundation (CNS-1725797) and administered by the Center for Scientific Computing (CSC). The CSC is supported by the California NanoSystems Institute and the Materials Research Science and Engineering Center (MRSEC; NSF DMR 1720256) at UC Santa Barbara.

\end{acknowledgements}

\software{%
\texttt{MESA} \citep[v15140,v21.12.1,v22.05.1;][]{MESAI,MESAII,MESAIII,MESAIV,MESAV,MESAVI}, 
\texttt{py\_mesa\_reader} \citep{bill_wolf_2017_826958},
\texttt{ipython/jupyter} \citep{perez_2007_aa,kluyver_2016_aa},
\texttt{matplotlib} \citep{hunter_2007_aa},
\texttt{NumPy} \citep{numpy2020}, 
\texttt{SciPy} \citep{scipy2020}, 
\texttt{Astropy} \citep{astropy:2013,astropy:2018},
and 
\texttt{Python} from \href{https://www.python.org}{python.org}
.}



\bibliography{flash,mesa,sofeware}{}
\bibliographystyle{aasjournal}

\end{document}